\documentstyle[preprint,prl,aps]{revtex}
\begin{document}
\draft
%
%
\title{Why we don't see the Schr\"{o}dinger's cat state?}
%
%
\author{Toshifumi Sakaguchi}
%
%
\address{
ASCII Corporation \\
Yoyogi 4-33-10, Shibuya-Ku, Tokyo, Japan
}
\date{\today}
\maketitle
%
%
\begin{abstract}
Schr\"{o}dinger's cat puzzle is resolved.  The reason why we do not see
a macroscopic superposition of states is cleared in the light of
Everett's formulation of quantum mechanics.
\end{abstract}
%
%
\narrowtext
%
%
Schr\"{o}dinger argued \cite{1,2} incompleteness of the Copenhagen
interpretation of quantum mechanics by showing a paradoxical thought
experiment known as ``Schr\"{o}dinger's cat'':  A cat is placed in a
sealed box with a device which releases a fatal dose of cyanide if a
radioactive decay is detected.  After a while, a human opens the box to
see if the cat is alive or dead.  According to the Copenhagen
interpretation, the cat was neither alive nor dead until the box was
opened, and when the human peeked in the box the wavefunction of the
cat collapsed into one of the two alternatives (otherwise the human
might be able to see a live cat after having seen a dead cat!). The
paradox is that the cat presumably knew if it was alive {\it before\/}
the box was opened, that is, the cat must have been either alive or
dead {\it before\/} the human opened the box.

As long as we adhere to the Copenhagen interpretation, the paradox will
not be resolved scientifically, because it provides no means to
analyze the whole system including the observer system.  In the
Copenhagen interpretation, the measurement process cannot be described
by any equation, but instead must be implemented by hand.

Everett proposed a new formulation of quantum mechanics \cite{3}, in
which the collapsing process of the wavefunction is ascribed to a
branching process of the observer state. That is, the wavefunction of
the cat never collapses into one of the two alternatives but instead
the wavefunction of the human branches into two states, each of which
being correlated to one of the alternatives.  One then may ask why the
observer state always correlates to the live-cat state and the dead-cat
state, rather than the other possible superposed alternatives, or to the
``Schr\"{o}dinger's cat state''.  Recent studies
\cite{4,5,6,7,8,9,10,11,12,13} have been focused on this issue and people
have tried to show how the preferred basis can be selected through the
process of interaction with the environment.

In this Letter, I will give an answer to this puzzle from a different
point of view.  In order to understand what it means, we need to refine
the formulation.  The answer will then follow as a natural consequence.

In Everett's formulation of quantum mechanics, we do {\it not\/}
interpret the squared modulus of the wavefunction as the probability,
but interpret it as a {\it measure\/} which satisfies the following
property:  When a subset of a superposition
$\sum_{i}X_{i}|\xi^{i}\rangle$
is described as a single element $Y|\eta\rangle$;
\begin{eqnarray}
Y|\eta \rangle &=& \sum_{i \in {\cal I}}X_{i}|\xi^{i}\rangle,
\label{a1}
\end{eqnarray}
then the corresponding measure $m$ must satisfy the relation;
\begin{eqnarray}
m(Y) &=& \sum_{i \in {\cal I}} m(X_{i}).
\label{a2}
\end{eqnarray}
In Eq.\ (\ref{a1}), the elements in the set
$\{|\xi^{i}\rangle\}$
have no overlap with each other, that is,
$\langle \xi^{i} | \xi^{j} \rangle = 0$ if $i \neq j$.
To distinguish the coefficients from the states uniquely, we
require that the states themselves are always normalized and that the
measure is a function of the absolute value of the coefficients.  Then
the function $m$ can be determined uniquely \cite{3}:
\begin{eqnarray}
m(|X_{i}|) = |X_{i}|^{2}.
\label{a3}
\end{eqnarray}
This is all about Everett's formulation of quantum mechanics.
It has nothing to do with the probability.

On the other hand, {\it if one is interested in the measurement process\/},
the conventional probabilistic interpretation can be {\it derived\/} as
follows.

We describe an observer state in which an
observed value $\alpha^{i}$, one of the eigenvalues of an observable
$\alpha$, is recoded, as $|[\alpha^{i}]\rangle$.  When we observe an
object system prepared in an {\it eigenstate\/} $|\alpha^{i}\rangle$,
we always get the value $\alpha^{i}$. Therefore, the measurement process
will be described:
\begin{eqnarray}
U(t) |\alpha^{i} \rangle|[\:] \rangle = |\alpha^{i} \rangle|[\alpha^{i}]
\rangle,
\label{a4}
\end{eqnarray}
where $U(t)$ is a time evolution operator obtained from a Hamiltonian
which includes the interaction between the object and observer systems.

We now prepare $N$ identical object systems in a same state
$| \psi \rangle = \sum_{i=1}^{M} C_{i} | \alpha^{i} \rangle$,
and observe $\alpha$ for each object sequentially.  Applying
Eq.\ (\ref{a4}) to each measurement process sequentially, we will get a
{\it superposition\/} of branched states of the form;
\begin{eqnarray}
C_{p} C_{q} \ldots C_{r} | \alpha^{p} \rangle| \alpha^{q} \rangle \ldots
|\alpha^{r} \rangle| [ \alpha^{p} \alpha^{q} \ldots \alpha^{r} ] \rangle.
\label{a5}
\end{eqnarray}
Measure $m(n_{1}, n_{2}, \ldots, n_{M})$ assigned to a subset
of the superposition of the branched states where
$\alpha^{1}$ is recorded $n_{1}$ times,
$\alpha^{2}$ is recorded $n_{2}$ times,
$\ldots$, and
$\alpha^{M}$ is recorded $n_{M}$ times
in each memory $[ \ldots ]$ is calculated as follows.  Each state vector
described above has the measure
$|C_{1}|^{2n_{1}}|C_{2}|^{2n_{2}} \ldots |C_{M}|^{2n_{M}}$
and there are
$N! / \prod_{i = 1}^{M} n_{i}!$
states in the subset, hence we get,
\begin{eqnarray}
m(n_{1}, n_{2}, \ldots, n_{M}) =
\frac{N!} { \prod_{i = 1}^{M} n_{i}! }
\prod_{i = 1}^{M} | C_{i} |^{2n_{i} },
\label{a6}
\end{eqnarray}
where
\begin{eqnarray}
N &=& \sum_{i = 1}^{M} n_{i}.
\label{a7}
\end{eqnarray}

A subset which gives a maximum value of the measure
$m(n_{1}, n_{2}, \ldots, n_{M})$
in the limit
$N \rightarrow \infty$
can be obtained from the following equations:
\begin{eqnarray}
F &=& \ln m( n_{1}, n_{2}, \ldots, n_{M} )
- \lambda (N - \sum_{i = 1}^{M} n_{i}) \nonumber \\
&\simeq&
N \ln N - N - \sum_{i = 1}^{M}
(n_{i} \ln n_{i} - n_{i})
+ \sum_{i = 1}^{M} n_{i} \ln | C_{i} |^{2} \nonumber \\
&-& \lambda (N - \sum_{i = 1}^{M} n_{i}),
\label{a8}
\end{eqnarray}
and
\begin{eqnarray}
\frac{\partial F}{\partial n_{i}} &=& 0 \hspace{1cm} (i = 1, 2, \ldots, M),
\label{a9} \\
\frac{\partial F}{\partial \lambda} &=& 0.
\label{a10}
\end{eqnarray}
{}From Eq.\ (\ref{a9}) we get
$n_{i} = e^{\lambda} |C_{i}|^{2}$,
and Eq.\ (\ref{a10}) yields
$N = e^{\lambda} \sum_{i = 1}^{M} |C_{i}|^{2} = e^{\lambda}$,
hence we obtain,
\begin{eqnarray}
\frac{n_{i}}{N} &=& |C_{i}|^{2}.
\label{a11}
\end{eqnarray}
In the limit
$N \rightarrow \infty$,
the measure $R_{N}$ assigned to the
remaining subset, to which the above elements do not belong,
vanishes:
\begin{eqnarray}
\lim_{N \rightarrow \infty} R_{N} &=&
\lim_{N \rightarrow \infty}
\{1 - m(N|C_{1}|^{2}, N|C_{2}|^{2}, \ldots, N|C_{M}|^{2})\} \nonumber \\
&=& \lim_{N \rightarrow \infty}
\{1 -  \frac{N!}{\prod_{i = 1}^{M} n_{i}!}
\prod_{i = 1}^{M} ( \frac{n_{i}}{N} )^{n_{i}} \} \nonumber \\
&=& 0.
\label{a12}
\end{eqnarray}
Therefore, in almost all elements of the superposition, the observer
who is described in {\it each\/} element will conclude that the
eigenvalue $\alpha^{i}$ can be obtained with probability
$n_{i}/N = |C_{i}|^{2}$ {\it in his world\/}.

One may attempt to require the disappearance of coherence
\cite{14,15,16,17,18,19,20,21} among branched observer states, i.e.,
that the observer states {\it must\/} be orthogonal to each other.
However, the coherence cannot be recognized by the observer described
by each branched observer state in {\it principle\/}.  This is
guaranteed by the linearity of the time evolution operator.  Even
though there {\it is\/} a moment when the states are coherent because
of the {\it continuous\/} measurement process, we do not notice the
branching process.  Therefore, it is superfluous to add the extra
condition to complete the derivation.

We now give an answer to the Schr\"{o}dinger's cat puzzle.  Let
$|\gamma_{live}\rangle$ and $|\gamma_{dead}\rangle$ be the live-cat
state and the dead-cat state, respectively.  Note that these vectors are
not necessarily eigenvectors of an operator, but are two solutions of
Schr\"{o}dinger's equation.

When the cat is in the {\it definite\/} state $|\gamma_{live}\rangle$
{\it or\/} $|\gamma_{dead}\rangle$, rather than a superposition of the
two states, we always find the live cat {\it or\/} dead cat when we
peek in the box. That is, if the cat state is $|\gamma_{live}\rangle$
then we always see the live cat and if it is $|\gamma_{dead}\rangle$
then we always see the dead cat. Therefore, the observation process
will be described by;
\begin{eqnarray}
U(t) |\gamma_{live}\rangle |[\:]\rangle =
|\gamma_{live}\rangle |[\gamma_{live}]\rangle,
\label{1}
\end{eqnarray}
and
\begin{eqnarray}
U(t) |\gamma_{dead}\rangle |[\:]\rangle =
|\gamma_{dead}\rangle |[\gamma_{dead}]\rangle.
\label{2}
\end{eqnarray}
{}From Eqs.\ (\ref{1}) and (\ref{2}) and from the linearity of the
time evolution operator $U(t)$, it follows that
\begin{eqnarray}
&U(t)&(a |\gamma_{live}\rangle + b |\gamma_{dead}\rangle)
|[\:]\rangle \nonumber \\
&=&
a |\gamma_{live}\rangle |[\gamma_{live}]\rangle +
b |\gamma_{dead}\rangle |[\gamma_{dead}]\rangle,
\label{3}
\end{eqnarray}
where $a$ and $b$ are non-zero complex numbers.
It means that when the cat is in a superposition;
\begin{eqnarray}
a |\gamma_{live}\rangle + b |\gamma_{dead}\rangle,
\label{4}
\end{eqnarray}
and we look in the box, {\it our state\/} $|[\:]\rangle$ branches into
two states $|[\gamma_{live}]\rangle$ and $|[\gamma_{dead}]\rangle$.
Note that the observer states in Eqs.\ (\ref{1}) and (\ref{2}), where
there is no other alternative, and those in Eq.\ (\ref{3}), which
appear in the superposition, are {\it exactly\/} the same.  This is the
reason why we cannot recognize the branching process.  One may try to
investigate the details of the observer states to see if the observer
states can be considered as representing ourselves, or if the observer
described by the observer states can be thought of as having a
consciousness.  However, it will be neither practicable nor indispensable.
We only have to know that the observer states which correspond to
``observer sees live cat'' and ``observer sees dead cat'' are
characterized by Eqs.\ (\ref{1}) and (\ref{2}), respectively. It should
be noted that the {\it measure\/} assigned to each observer state in
Eq.\ (\ref{3}) is invariant under any basis transformations.

On the other hand, if we can see the superposition of states {\it as it
is\/}, the observation process must be described:
\begin{eqnarray}
&U^{\prime}(t)&(a |\gamma_{live}\rangle
+ b |\gamma_{dead}\rangle) |[\:]\rangle \nonumber \\
&=& (a |\gamma_{live}\rangle + b |\gamma_{dead}\rangle) |[\Phi_{+}]\rangle,
\label{5}
\end{eqnarray}
and
\begin{eqnarray}
&U^{\prime}(t)&(a |\gamma_{live}\rangle
- b |\gamma_{dead}\rangle) |[\:]\rangle \nonumber \\
&=& (a |\gamma_{live}\rangle - b |\gamma_{dead}\rangle) |[\Phi_{-}]\rangle.
\label{6}
\end{eqnarray}
{}From Eqs.\ (\ref{5}) and (\ref{6}) and from the linearity of the
time evolution operator $U^{\prime}(t)$, it follows that
\begin{eqnarray}
U^{\prime}(t)|\gamma_{live}\rangle |[\:]\rangle =
&\frac{1}{2a}& \{
(a |\gamma_{live}\rangle + b |\gamma_{dead}\rangle) |[\Phi_{+}]\rangle
\nonumber \\
&+& (a |\gamma_{live}\rangle - b |\gamma_{dead}\rangle) |[\Phi_{-}]\rangle
\},
\label{7}
\end{eqnarray}
and
\begin{eqnarray}
U^{\prime}(t)|\gamma_{dead}\rangle |[\:]\rangle =
&\frac{1}{2b}& \{
(a |\gamma_{live}\rangle + b |\gamma_{dead}\rangle) |[\Phi_{+}]\rangle
\nonumber \\
&-& (a |\gamma_{live}\rangle - b |\gamma_{dead}\rangle) |[\Phi_{-}]\rangle
\}.
\label{8}
\end{eqnarray}
Eqs.\ (\ref{7}) and (\ref{8}) show that we cannot see the live cat or
the dead cat {\it even when the cat is in the definite state
$|\gamma_{live}\rangle$ or $|\gamma_{dead}\rangle$ \/}.  It will be
possible to realize such an interaction only in {\it principle\/}.

The situation described above demonstrates Bohr's complementarity
\cite{22}, which was {\it derived\/} as a theorem in another paper
\cite{23}.
%
%

%
%
%
\end{document}